\documentclass{article}

\PassOptionsToPackage{numbers,compress}{natbib}

\usepackage[preprint]{neurips_2025}

\usepackage[utf8]{inputenc} 
\usepackage[T1]{fontenc}    
\usepackage{url}            
\usepackage{booktabs}       
\usepackage{amsfonts}       
\usepackage{nicefrac}       
\usepackage{microtype}      
\usepackage{xcolor}         
\usepackage[pdftex]{graphicx} 
\usepackage{algorithm}      
\usepackage{algpseudocode}  
\usepackage{subcaption}     
\usepackage{multirow}       

\usepackage[colorlinks=true,linkcolor=blue,citecolor=blue,urlcolor=blue]{hyperref}

\title{Beyond the Buzz: \\A Pragmatic Take on Inference Disaggregation}

\author{%
  Tiyasa Mitra \\
  \And
  Ritika Borkar \\
  \And
  Nidhi Bhatia \\
  \And
  Ramon Matas \\
  \And 
  Shivam Raj \\
  \And
  Dheevatsa Mudigere \\
  \And
  Ritchie Zhao \\
  \And
  Maximilian Golub \\
  \And
  Arpan Dutta \\
  \And
  Sailaja Madduri \\
  \And
  Dharmesh Jani \\
  \And
  Brian Pharris \\
  \And
  Bita Darvish Rouhani \\
  \AND
  \textnormal{NVIDIA Corporation}
}

\begin{document}

\maketitle

\begin{abstract}
  As inference scales to multi-node deployments, disaggregation—splitting inference into distinct phases—offers a promising path to 
  improving the throughput-interactivity Pareto frontier. Despite growing enthusiasm and a surge of open-source efforts, practical 
  deployment of disaggregated serving remains limited due to the complexity of the optimization search space and system-level coordination. 
  In this paper, we present the first systematic study of disaggregated inference at scale, evaluating hundreds of thousands of design points across diverse workloads and hardware configurations. We find that disaggregation is most effective for prefill-heavy traffic patterns and larger models. Our results highlight the critical role of dynamic rate matching and elastic scaling in achieving Pareto-optimal performance. Our findings offer actionable insights for efficient disaggregated deployments to navigate the trade-off between system throughput and interactivity.
\end{abstract} 
\section{Introduction}
One of the primary drivers of modern AI applications is scale. Inference serving is rapidly evolving from traditional single-node endpoints to multi-node deployments at data center scale. This shift unlocks new opportunities for system-level optimization, enabling exploration across a broader, more expressive design space.

The community is actively exploring a range of techniques to improve the Pareto frontier of server throughput (amortized cost) and interactivity (quality of service) for inference serving. Disaggregation is a prominent example, offering the potential to expand the throughput–interactivity trade-off when applied effectively. Reflecting this promise, the past year has seen a surge in research efforts and open-source implementations. Despite growing interest in disaggregation, adoption at scale has been limited--–mainly due to the complexity of the underlying design space.

At its core, disaggregation is straightforward: divide inference into phases with distinct compute characteristics and optimize each phase independently. In autoregressive LLMs, this typically corresponds to separating the prefill and decode phases. This paper focuses on LLMs as the primary case study due to their widespread use, though the same principles should apply to VLMs.

The design space for disaggregation is broad and highly dependent on traffic characteristics. Unlocking performance gains through disaggregation requires careful decisions around model sharding, concurrency, and rate matching between phases to balance throughput and interactivity trade-offs. In this work, we systematically quantify the search space for LLM disaggregation and examine the system-level trade-offs and implications within this emerging landscape.

\begin{figure}[tp]
    \centering
    \includegraphics[width=0.95\columnwidth]{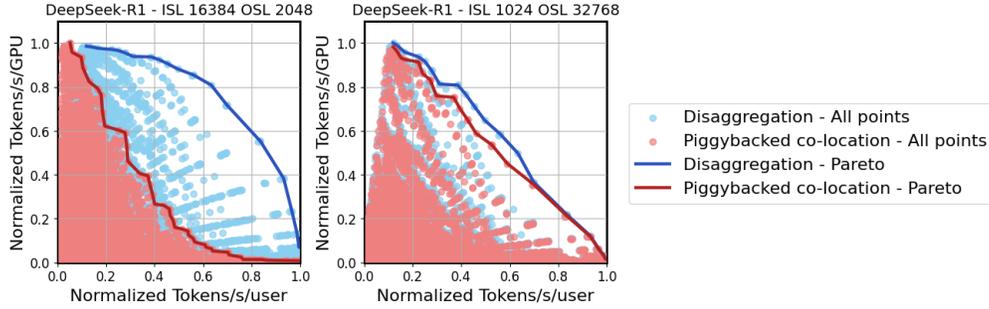}
    \caption{Throughput–interactivity Pareto frontier for DeepSeek-R1. 
    The benefits of disaggregated serving vary with the target tokens/s/user (i.e., interactivity) and 
    traffic patterns: (left) prefill-heavy vs. (right) generation-heavy traffic. Most results in this paper are presented in normalized form, as our primary objective is to convey trends rather than make specific performance claims.}
    \label{fig:overview}
\end{figure}

This paper seeks to clarify the practical aspects of disaggregated serving and to provide design guidance for large-scale deployments. While disaggregation holds significant promise for improving LLM inference, it also comes with new optimization challenges. Figure \ref{fig:overview} provides an example of how the benefits of disaggregation, over piggybacked co-location, can vary significantly across different traffic patterns – i.e., Input Sequence Length (ISL) and Output Sequence Length (OSL). 

This paper, for the first time, extensively explores the design space of large-scale disaggregated inference by simulating hundreds of thousands of design points across a range of workloads, traffic patterns, and hardware configurations. Our analysis reveals that disaggregation provides the greatest benefits in prefill-heavy traffic scenarios (i.e., ISL $>>$ OSL) and when serving larger models (e.g., >$10B$ parameters). Our results also underscore the importance of dynamic rate matching and elastic scaling to fully realize the advantages of disaggregation. Similarly, for co-located serving, we find that the effectiveness of context chunking is highly sensitive to the attention mechanism (e.g., Multi-Latent Attention (MLA) vs. Group Query Attention (GQA)) and is most beneficial under relaxed latency targets and generation-heavy traffic patterns. Through extensive experimentation and system-level analysis, we hope this paper serves as a foundation for building high-performance disaggregated inference systems at scale in the future.

\section{Background}
In traditional co-located LLM inference serving, both the prefill and decode phases occur on the same model instance within a monolithic pipeline. To maximize throughput, requests in different phases are batched together to share model weights and GPU memory. Many deployments employ in-flight batching (IFB), which allows new requests to be added to a batch as soon as an in-flight request completes. Piggybacking \citep{sarathi2023,agrawal2024} builds upon IFB by reducing decode stalls through context chunking when new requests are introduced.

Despite these optimizations, co-located serving forces a single model instance to simultaneously optimize for two metrics: low First Token Latency (FTL) for new prompts and low Token-to-Token Latency (TTL) for ongoing generation. Each metric exhibits different bottlenecks, leading to inherent tension in resource scheduling.

\begin{figure}[htbp]
    \centering
    \includegraphics[width=1.0\columnwidth]{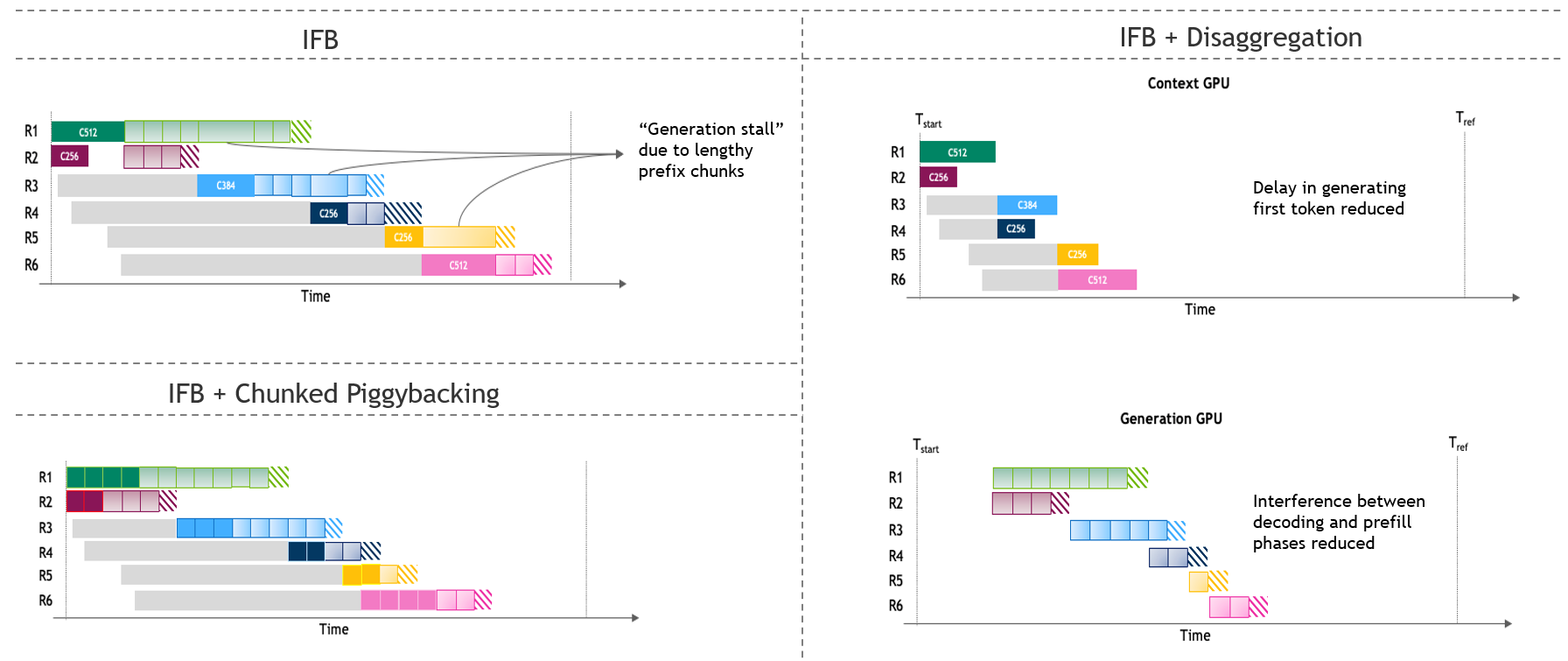}
    \caption{Visualization of (left) co-located and (right) disaggregated inference serving, illustrating the temporal distribution of prefill (dark boxes) and decode iterations (light boxes) across different requests (color-coded).}
    \label{fig:figure2}
\end{figure}

In contrast, disaggregated inference serving \citep{tensorrtllm2024,vllm2024,distserve2024,mooncake2025,pdserve2024} decouples the prefill and decode phases, allowing each to run on a separate model instance — potentially across different GPUs. This separation enables each phase to independently adopt model partitioning and batching strategies tailored to its performance targets. Moreover, it eliminates artificial slowdowns in prefill caused by strict TTL service-level agreements, as seen in piggybacking. Figure \ref{fig:figure2} illustrates these modes of serving.

As demonstrated in Figure \ref{fig:overview}, disaggregation is not a universal solution. In the following sections, we examine the performance benefits of disaggregated inference serving across a broad design space. 
Appendix \ref{appendixA} summarizes the key metrics referenced throughout the paper.

\section{Design space exploration}
A versatile inference optimization should maximize the area under the throughput–interactivity Pareto frontier (see Figure~\ref{fig:overview} as an example). In the case of disaggregated serving, achieving this goal requires optimization across two distinct dimensions: \textbf{(i)} the model partitioning strategy for prefill (a.k.a., context) and decode (a.k.a., generation) instances, and \textbf{(ii)} the scaling/rate matching strategy for prefill and decode GPUs.

\subsection{Model partitioning}
Each point along the Pareto frontier corresponds to a different model partitioning strategy. To explore this space, we evaluate different parallelism strategies including Tensor Parallelism (TP) \citep{shoeybi2020}, Expert Parallelism (EP) \citep{gshard2021}, Pipeline Parallelism (PP) \citep{gpipe2019,pipedream2019}, Chunked Pipeline Parallelism (CPP), and TEP (Tensor Parallel Attention and EP FFNs), across a wide range of batch sizes. 

The optimal partitioning strategy for a given model depends on several factors: the serving mode (e.g., co-located or disaggregated), traffic characteristics (input sequence length, output sequence length, and queries per second), target hardware, and latency constraints. Throughout this paper, we use a proprietary, high-fidelity GPU performance simulator designed for datacenter-scale inference. The simulator takes as input the model architecture, traffic pattern, and GPU configuration, and outputs the corresponding latency and throughput across different batch sizes and parallelism strategies. These outputs are used to construct Pareto frontiers under various serving conditions.

For co-located serving, we evaluate configurations both with and without context-chunked piggybacking and explore the sensitivity of piggybacking to model architecture and traffic patterns. In piggybacked setups, our simulator determines the optimal mix of prefill and decode tokens in a batch for each ISL–OSL combination. This ratio varies across the Pareto frontier and depends on the latency constraints.

In the disaggregated setting, a key source of performance gain is the ability to use different model partitioning strategies for the prefill and decode stages. To reflect this, we simulate the prefill and decode pools separately, allowing each to independently optimize for its corresponding service level agreements.
Our analysis focuses on modern Blackwell systems \citep{blackwell2024} using FP4 precision \citep{microscaling2023}, which represent the state of the art in LLM inference infrastructure.

\subsection{Scaling and rate matching}
\label{sec:scaling_strategy}
Once the optimal model partitioning for prefill and decode is identified, disaggregated serving requires a rate matching strategy to determine the 
appropriate ratio of prefill to decode instances and ensure a balanced throughput between the two phases.

To construct the Pareto frontier shown in Figure \ref{fig:overview}, we first fix the prefill mapping that satisfies the FTL constraint. 
Then, for each candidate decode mapping, we use a rate matching algorithm that employs an integer solver to find the right balance between 
the throughput of prefill and decode phases – subject to the TTL constraint and total GPU count minimization. Details can be found in Appendix \ref{appendixB}. All design points with an FTL 
 $>10$ seconds, a relaxed yet practical constraint, are excluded from our search space.

\begin{figure}[htbp]
    \centering
    \includegraphics[width=0.95\textwidth]{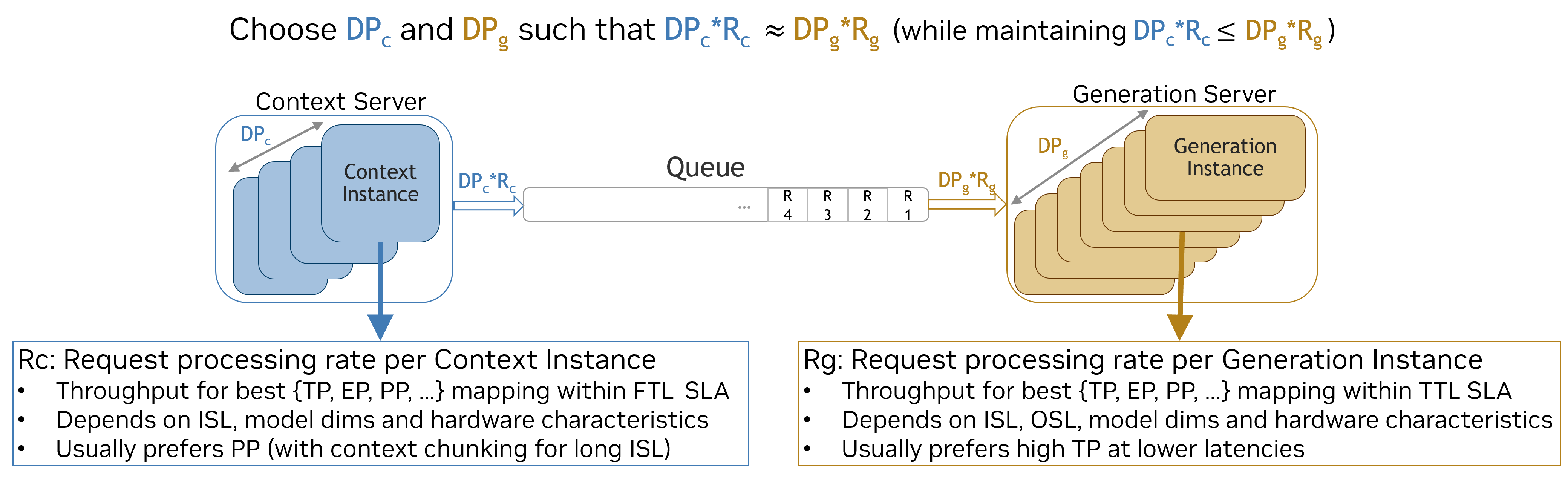}
    \caption{High-level overview of rate matching for disaggregated serving. KV cache and weights are hosted in HBM memory and capacity constraints are accounted for.}
    \label{fig:rate_matching}
\end{figure}

Figure~\ref{fig:rate_matching} shows the high-level overview of the rate matching methodology used to quantify each disaggregated design point. 
Each blue circle in Figure~\ref{fig:overview} is the output of a rate matching step. 
Our simulation assumes a datacenter setting with sufficient GPUs and incoming requests to fully utilize the rate-matched deployment.
We further assume the KV cache produced at each layer by the prefill pool is transferred to the generation pool immediately as it becomes available, overlapping with the computation of subsequent layers. The implications of this assumption are discussed in Section~\ref{kv_bw_requirements}.

\section{Disaggregation in practice}
The performance gains from disaggregated serving depend on multiple factors—including target latency, underlying model architecture and scale, traffic patterns, and hardware configurations. In this section, we analyze the sensitivity of disaggregated serving to each of these dimensions.

In real-world deployments, service level agreements (SLA) are typically defined by two latency metrics: (i) FTL, ranging from hundreds of milliseconds to several minutes, and (ii) TTL, typically spanning a few milliseconds. The reciprocal of TTL (i.e., $1/$TTL) serves as a proxy for interactivity, measured in tokens per second per user. Together, FTL and TTL determine where a system should operate along the throughput–interactivity Pareto frontier for optimal performance.

\begin{figure}[htbp]
    \centering
    \includegraphics[width=0.8\columnwidth]{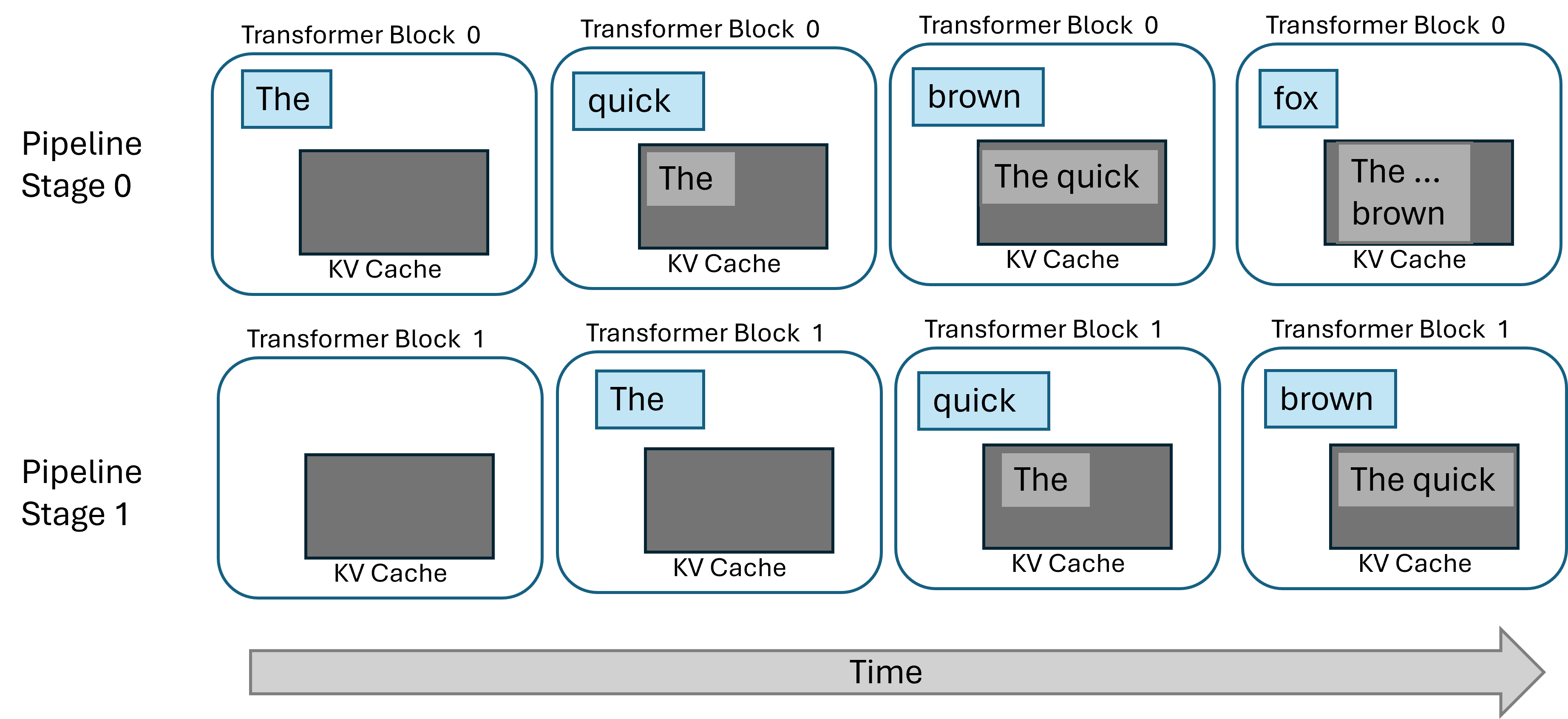}
    \caption{High-level overview of chunked pipeline parallelism. It works by: (i) splitting the input sequence into smaller chunks, (ii) processing each chunk independently, using the KV cache from previous chunks but not their outputs, and (iii) overlapping the processing of earlier layers of new chunks with the later layers of previous ones using pipeline parallelism.}
    \label{fig:chunked_pipelining}
\end{figure}

\begin{figure}[htbp]
    \centering
    \includegraphics[width=0.7\columnwidth]{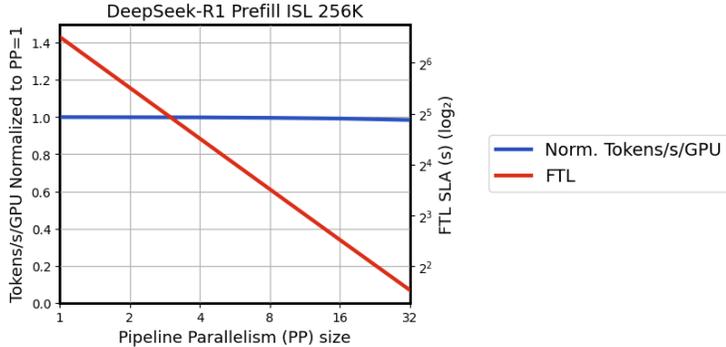}
    \caption{Chunked pipeline parallelism during Prefill is an optimal strategy to maximize throughput while complying with strict FTL SLA. Prefill performance is shown for DeepSeek-R1 with ISL of 256K on 64 GPUs using EP and PP (EP $\times$ PP = 64).}
    \label{fig:ctx_pp}
\end{figure}

In disaggregated serving, FTL constraints apply only to the prefill (context) pool. To maintain high system throughput while reducing FTL across mixed-length sequences, we found Chunked Pipeline Parallelism (CPP) to be especially effective. As shown in Figure \ref{fig:chunked_pipelining}, chunked pipelining splits context processing into smaller, parallel segments. This allows context GPUs to handle long sequences within the given FTL, without the complexity of wide tensor parallelism. Figure ~\ref{fig:ctx_pp} shows, for DeepSeek-R1, how FTL can be reduced as we increase the PP, while keeping throughput high.

TTL constraints, on the other hand, govern the decode (generation) pool. In Figure \ref{fig:overview}, moving left to right corresponds to increasingly stringent TTL requirements (i.e., higher tokens/s/user). As TTL constraints tighten, configurations shift toward smaller batch sizes and greater tensor parallelism.

Consider DeepSeek-R1 \citep{deepseekr12025} with ISL of 16k and OSL of 2k: across the Pareto frontier, expert parallelism within the NVLink domain is consistently preferred. However, attention computation transitions from data parallelism in the high-throughput regime to tensor parallelism under tighter TTL constraints. Batch sizes are in the hundreds at the high-throughput end, decreasing progressively as interactivity increases.
A similar trend is observed in Llama-3.1-70B \citep{llama2024}, where tensor parallelism scales from 2$\times$ to 64$\times$ as TTL constraints tighten, with batching behavior closely mirroring that of DeepSeek-R1.
While both co-located and disaggregated serving favor high tensor parallelism under tight TTL SLAs, disaggregated decoding is able to pursue this strategy more aggressively. Freed from the need to balance math-heavy prefill performance with decoding speed, disaggregated decode pools can better adapt to tightening latency demands — leading to superior performance in the medium-latency regime.

\subsection{Model sensitivity}
In this section, we examine the sensitivity of disaggregated serving to model architecture and size.

\begin{figure}[htbp]
    \centering
    \includegraphics[width=0.9\columnwidth]{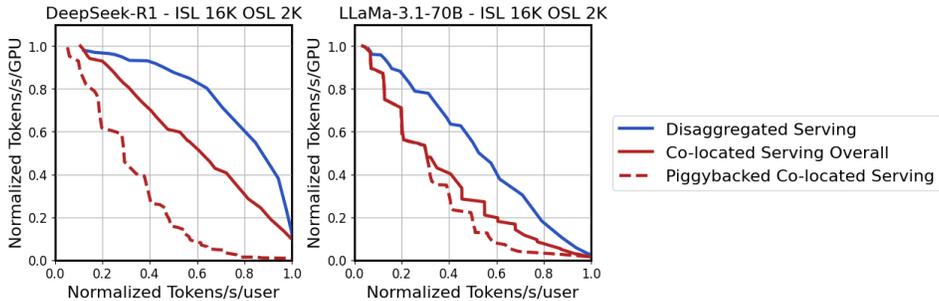}
    \caption{Disaggregated vs. co-located serving. Co-located serving overall (red-solid) is the superposition of piggybacked (red-dotted) and non-piggybacked configurations.} 
    \label{fig:disagg_model_arch}
\end{figure}

\textbf{Model architecture sensitivity.}
Model architecture plays a key role in inference serving decisions. Figure~\ref{fig:disagg_model_arch} compares disaggregated and co-located serving under context-heavy traffic for DeepSeek-R1 and Llama-3.1-70B. We evaluate the sensitivity of disaggregation to traffic pattern in the next section.
Note that the benefits of disaggregation manifest differently across different latency regimes for each model architecture. 
Our analysis further reveals that DeepSeek-R1 experiences additional overhead in piggybacked co-located serving due to prefill chunking—specifically, redundant computation of down and up projections in multi-latent attention for each prefill chunk. This can be mitigated by temporarily caching the up-projected KV values from earlier chunks. To capture these trade-offs, our co-located baseline Pareto curves include both piggybacked and non-piggybacked configurations.

\begin{figure}[htbp]
    \centering
    \includegraphics[width=0.9\columnwidth]{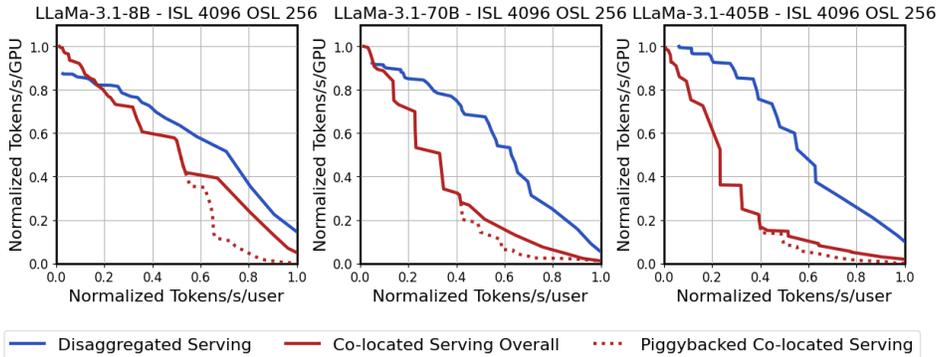}
    \caption{Larger models benefit more from disaggregated serving due to a richer search space.}
    \label{fig:disagg_model_size}
\end{figure}

\textbf{Model size sensitivity.} 
Figure~\ref{fig:disagg_model_size} presents the throughput-latency characteristics for Llama 8B, 70B, and 405B under both disaggregated and co-located serving configurations. Our analysis indicates that the benefits of disaggregated inferencing become more pronounced with larger models. This enhanced performance can be attributed to the fact that larger models are typically mapped across more GPUs, enabling a broader range of parallelization strategies. Consequently, the advantage of selecting distinct model mappings for prefill and decode phases becomes more significant.

\subsection{Traffic sensitivity}
\begin{figure}[htbp]
    \centering
    \includegraphics[width=1.0\columnwidth]{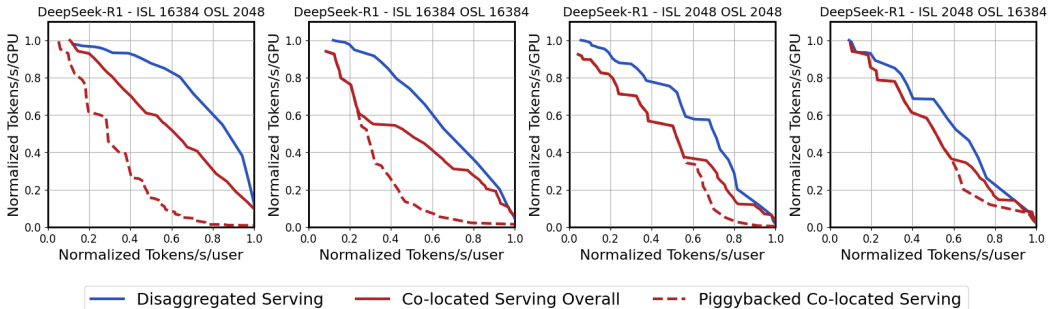}
    \caption{Disaggregation helps most under prefill-heavy traffic.}
    \label{fig:traffic_sensitivity} 
\end{figure}

A critical factor in estimating the performance of disaggregated serving is the traffic pattern. 
In Figure \ref{fig:traffic_sensitivity}, we present the Pareto for DeepSeek-R1 across four distinct traffic patterns. Our analysis reveals that the benefits of disaggregation are most pronounced for prefill-heavy workloads where mappings, if prioritized to balance decoding speed, can significantly compromise prefill processing throughput. Similarly, for co-located serving, piggybacking is most promising on decode-heavy traffic.

It is important to note that our simulation with constant ISL and OSL represents an approximation of dynamic traffic, where these values correspond to power-of-two approximations of the 50th percentile ISL and OSL.
See Appendix \ref{appendixD} for a demonstration of how using the P50 ISL/OSL provides a reliable representation of the Pareto frontier under dynamic real-world traffic conditions.

\subsection{Dynamic rate matching considerations}
\begin{figure}[htbp]
    \centering
    \includegraphics[width=0.5\columnwidth]{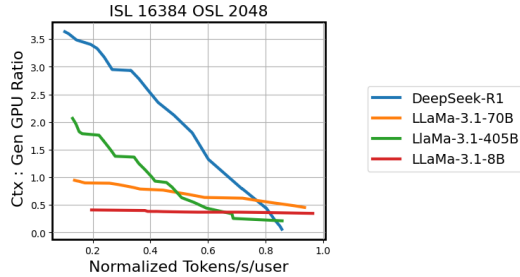}
    \caption{The optimal ratio of ctx-to-gen GPUs varies across models and target latencies.}
    \label{fig:figure6_ctx_gen_ratio}
\end{figure}

The optimal context-to-generation GPU ratio exhibits significant variation with model characteristics and target latency, 
as shown in Figure \ref{fig:figure6_ctx_gen_ratio}. Therefore, a versatile disaggregated serving system should incorporate a dynamic rate matching mechanism 
to adapt to changes in serving requirements. 

\begin{figure}[htbp]
    \centering
    \includegraphics[width=0.75\columnwidth]{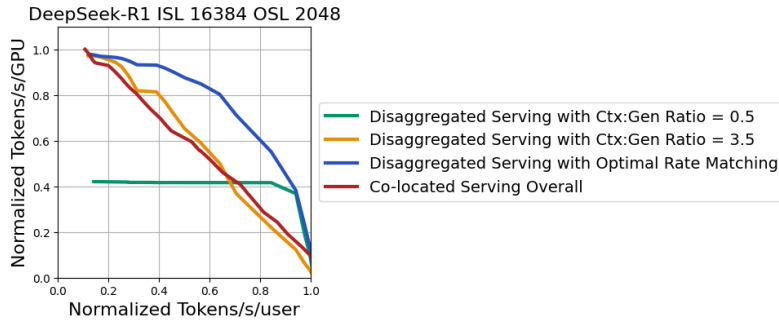}
    \caption{Optimal rate matching dynamically adapts Ctx:Gen ratio to deliver Pareto optimal performance. A ratio of 3.5 is performant at the most relaxed latency target but degrades as latency tightens. Conversely, a ratio of 0.5 favors tight latency but suffers significantly under relaxed latency.}
    \label{fig:fixed_ratios}
\end{figure}

Figure ~\ref{fig:fixed_ratios} illustrates the performance degradation of DeepSeek-R1 when rate matching is constrained to a fixed ratio. A similar effect is expected in small-scale GPU deployments, where limited resources can restrict the rate matching search space.

\subsection{NVLink sensitivity}

\begin{figure}[h]
    \centering
    \includegraphics[width=0.75\columnwidth]{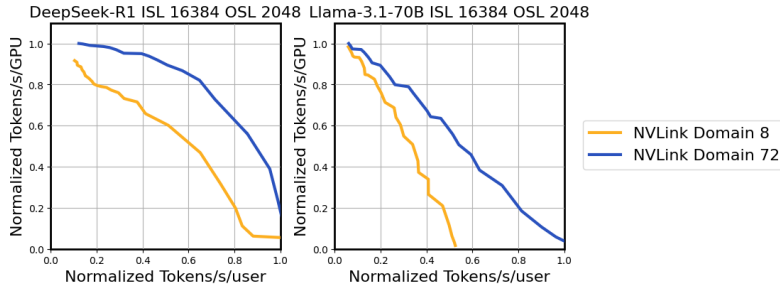}
    \caption{Larger NVLink domain helps disaggregated serving performance. DeepSeek-R1 benefits from higher EP and batching at medium-latency. Llama-3.1-70B benefits from high TP at low-latency.}
    \label{fig:nvlink_sensitivity}
\end{figure}

Figure \ref{fig:nvlink_sensitivity} shows the Pareto performance of disaggregated serving for two NVLink domain sizes. The analysis reveals that larger NVLink domains consistently enhance disaggregated serving performance. The benefit comes from the flexibility to choose wider expert and tensor parallelism during generation.

\section{Deployment considerations}
Disaggregated inferencing incurs a one-time overhead per request for transferring KV cache from prefill to decode GPUs. In this section, we analytically quantify the bandwidth requirements necessary to prevent KV cache transfer from becoming a performance bottleneck.

\subsection{Bandwidth requirements for KV cache transfer}
\label{kv_bw_requirements}
Prefill GPUs generate KV cache on a layer-by-layer basis, creating an opportunity to overlap KV transfer with 
prefill computation. The required egress bandwidth per GPU to fully overlap KV cache transfer with prefill
compute can be derived as:
    \begin{equation}
    BW_{egress} = \frac{N_{layers} \times BS_{prefill} \times ISL \times d_{head} \times N_{kv_{heads}} \times bytes_{{element}}}{FTL \times NumGPU_{prefill}}
    \end{equation}
    where $N_{layers}$ represents the number of layers in the model, $BS_{prefill}$ denotes the batch size of the prefill instance, $ISL$ indicates the input sequence length, $d_{head}$ represents the attention head dimension, $N_{kv_{heads}}$ specifies the number of KV heads, $bytes_{{element}}$ indicates the number of KV cache bytes per token, $FTL$ represents the time required for prefill computation to complete, and $NumGPU_{prefill}$ denotes the GPU count in each prefill instance that uniquely shards the KV cache.

    The ingress bandwidth requirement for decode GPUs is constrained by decode computation time to receive KV cache. The bandwidth per decode GPU can be derived as:
    \begin{equation}
    BW_{ingress} = \frac{N_{layers} \times BS_{decode} \times ISL \times d_{head} \times N_{kv_{heads}} \times bytes_{element}}{TTL \times OSL \times NumGPU_{decode}}
    \end{equation}
    where $BS_{decode}$ represents the batch size of the decode instance, $TTL$ indicates the time required to decode each output token, $OSL$ denotes the output sequence length, and $NumGPU_{decode}$ specifies the number of GPUs in each decode instance that uniquely shards the KV cache.
\begin{figure}[h]
    \centering
    \includegraphics[width=0.6\columnwidth]{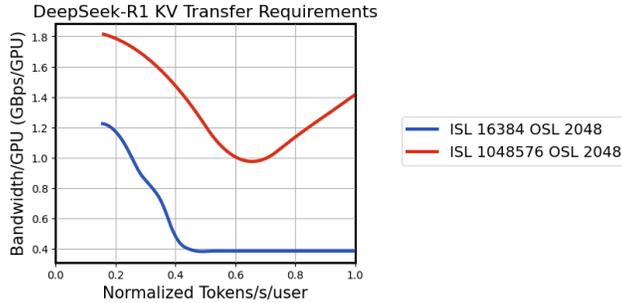}
    \caption{Bandwidth requirements for KV cache transfer: Maximum of egress and ingress bandwidth across various TTLs, showing the relationship between SLA constraints and bandwidth needs.}
    \label{fig:bandwidth_requirements}
\end{figure}

It's important to note that some parallelism schemes replicate the KV cache rather than sharding it. For example, when the tensor parallelism domain exceeds the number of KV heads, the KV cache is duplicated across tensor parallel ranks. The duplication factor in this case is equal to the ratio of tensor parallel ranks to KV heads.
As a result, when calculating per-GPU bandwidth requirements, only the GPUs that actually shard the KV cache should be considered in the normalization.

Due to the quadratic cost of attention during prefill, FTL scales superlinearly with ISL, whereas the KV cache size scales linearly. 
This divergence implies that the egress bandwidth requirement decreases as ISL increases.
On the decode side, both the KV cache size and TTL scale linearly with ISL, effectively canceling out their impact on ingress bandwidth. 
However, ingress bandwidth is inversely proportional to OSL. As TTL constraints tighten, more decode GPUs are required to meet latency targets, which effectively lowers the per-GPU ingress bandwidth requirement.

As for the impact of model scale on bandwidth requirements, FTL scales linearly with the number of active parameters. However, the KV cache size does not grow proportionally to the number of model parameters. Consequently, larger models with optimized attention (i.e., MLA in DeepSeek-R1) may require less egress bandwidth than smaller models with less efficient attention architectures.

Figure~\ref{fig:bandwidth_requirements} shows the maximum of egress and ingress bandwidth requirements for two sequence length combinations on DeepSeek-R1 under varying TTLs. 
Our analysis indicates that existing provisioned datacenter bandwidth is sufficient to support KV cache transfer without becoming a bottleneck.

\section{Related work}

Over the years, several works \citep{deepspeed2022, orca2022, pagedattention2023, scaling2023} have explored the efficiency of LLM serving at scale without reference to disaggregation. 
Recently, disaggregated serving has emerged as a new paradigm for improving the efficiency of large-scale LLM inference. In response to its potential, the past year has seen a wave of academic research \citep{distserve2024, splitwise2024, dejavu2024, dynaserve2025, kvdirect2025, tetriinfer2024, fastdecode2024, jiang2024hexgen, memserve2024, attentiondisagg2024} and implementations~\citep{tensorrtllm2024, vllm2024, mooncake2025, pdserve2024} exploring various facets of disaggregation. 
However, widespread adoption remains limited, mostly due to the inherent complexity of the design space—including challenges in workload balancing and parallelism.

While existing open-source implementations offer a valuable starting point, they fall short of providing concrete guidance on when and how disaggregation is beneficial. Similarly, prior research has largely focused on small-scale testbeds and peak throughput scenarios, without examining the full throughput–interactivity Pareto frontier.

To our knowledge, this work presents the first systematic study of disaggregated serving at datacenter scale, offering a comprehensive analysis of the key design trade-offs and practical considerations needed for real-world deployment.

\section{Future work}

The optimization space for large-scale inference serving is expanding rapidly with new algorithmic techniques and hardware innovations. Of the many interesting directions to explore, the impacts of KV cache reuse, speculation, inference-time compute techniques, and model architecture evolution appear to be particularly promising directions to pursue.

\section{Conclusions}
In this work, we present design principles for efficient disaggregated serving and evaluate its effectiveness at large scale. 
Our analysis shows that the optimal configurations for disaggregated serving depend on a combination of factors, including model size and architecture, traffic patterns, latency constraints, and hardware resources. We also highlight scenarios where disaggregation offers limited benefit—such as serving small-scale models or generation-heavy traffic.
These findings provide practical guidance for at scale deployment, where balancing throughput and latency remains a critical challenge in meeting the evolving demands of LLM serving.

{
\small
\bibliographystyle{unsrtnat}
\bibliography{references}
}

\newpage
\appendix
\appendix


\section{Terminology}
\label{appendixA}
This section defines key terms and metrics used throughout the paper. We use prefill and context processing interchangeably, as well as decode and generation.

\begin{table}[htbp]
    \caption{Summary of metrics used throughout the paper.}
    \label{tab:terminology}
    \centering
    \small
    \begin{tabular}{p{0.3\textwidth}p{0.1\textwidth}p{0.55\textwidth}}
        \toprule
        Metric & Acronym & Definition \\
        \midrule
        First Token Latency & FTL & Latency to run prefill and generate the first token. \\\\
        Token-to-Token Latency & TTL & Latency to generate each new token in decoding. \\ \\
        Tokens per Second per User & TPS & The rate of token generation per user ($1/$TTL). \\ \\
        Service Level Agreement & SLA & Agreed upon P50 TTL and FTL for a service. \\ \\
        Batch (a.k.a., concurrency) & – & The number of requests a system can serve per model replica. \\ \\
        Context Throughput (per GPU) & – & The throughput per context (prefill) GPU expressed in requests/second/GPU. This accounts for only the context (prefill) GPUs deployed. \\ \\
        Decode Throughput (per GPU) & – & The throughput per decode (generation) GPU expressed in tokens/second/GPU. This accounts for only the decode (generation) GPUs deployed. \\ \\
        Overall Throughput (per GPU) & – & The total throughput of the system expressed in tokens/second/GPU. This accounts for all GPUs deployed (prefill and decode). \\ \\
        Utilization & – & For a hardware resource, the percentage of execution time that the resource would be busy assuming 100\% efficiency. \\ \\
        \bottomrule
    \end{tabular}
\end{table}


\section{Procedure for optimizing prefill \& decode balance (a.k.a., rate matching)}
\label{appendixB}

\begin{algorithm}[htbp]
    \caption{Prefill Configuration Selection}
    \label{alg:prefill_config_selection}
    \begin{algorithmic}[1]
        \Procedure{PrefillConfigSelection~}{FTL\_cutoff, $tuple\_list$ (prefill\_config, FTL)}

            \State $best\_throughput \gets 0$
            \State $best\_config \gets \textrm{None}$

            \For {$(prefill\_config,\ FTL)$ in $tuple\_list$}
                \If {$FTL < FTL\_cutoff$}
                    \State $B \gets prefill\_config.batch\_size$
                    \State $G \gets prefill\_config.num\_gpus$
                    \State $throughput \gets \frac{B}{FTL \times G}$

                    \If {$throughput > best\_throughput$}
                        \State $best\_throughput \gets throughput$
                        \State $best\_config \gets prefill\_config$
                    \EndIf
                \EndIf
            \EndFor

            \State \Return $(best\_config,\ best\_throughput)$
        \EndProcedure
    \end{algorithmic}
\end{algorithm}

\begin{algorithm}[htbp]
    \caption{Rate Matching Prefill with Decode GPUs}
    \label{alg:rate_matching}
    \begin{algorithmic}[1]
        \Procedure{RateMatching~}{best\_prefill\_config, best\_prefill\_throughput, $decode\_tuple\_list$ (decode\_config, TTL), OSL, tolerance = 0.03}

            \State $rate\_matched\_list \gets [\,]$

            \For {$(decode\_config,\ TTL)$ in $decode\_tuple\_list$}
                \State $B \gets decode\_config.batch\_size$
                \State $G \gets decode\_config.num\_gpus$
                \State $decode\_throughput \gets \frac{B}{TTL \times G}$

                \State $decode\_request\_throughput \gets \frac{decode\_throughput}{OSL - 1}$
                \State $\alpha \gets \mathrm{round} \left( \frac{best\_prefill\_throughput}{decode\_request\_throughput},\ tolerance \right)$

                \State $num\_prefill\_gpus \gets \mathrm{numerator} (\alpha) \times G$
                \State $num\_decode\_gpus \gets \mathrm{denominator} (\alpha) \times best\_prefill\_config.num\_gpus$

                \State $throughput \gets \frac{decode\_throughput}{1 + \alpha}$

                \State $rate\_matched\_list.\mathrm{append} (throughput,\ num\_decode\_gpus,\ num\_prefill\_gpus)$
            \EndFor

            \State \Return $rate\_matched\_list$
        \EndProcedure   
    \end{algorithmic}
\end{algorithm}


\section{Using 50th percentile (P50) statistics as proxy for performance analysis}
\label{appendixD}

\begin{figure}[htbp]
    \centering
    \includegraphics[width=0.9\textwidth]{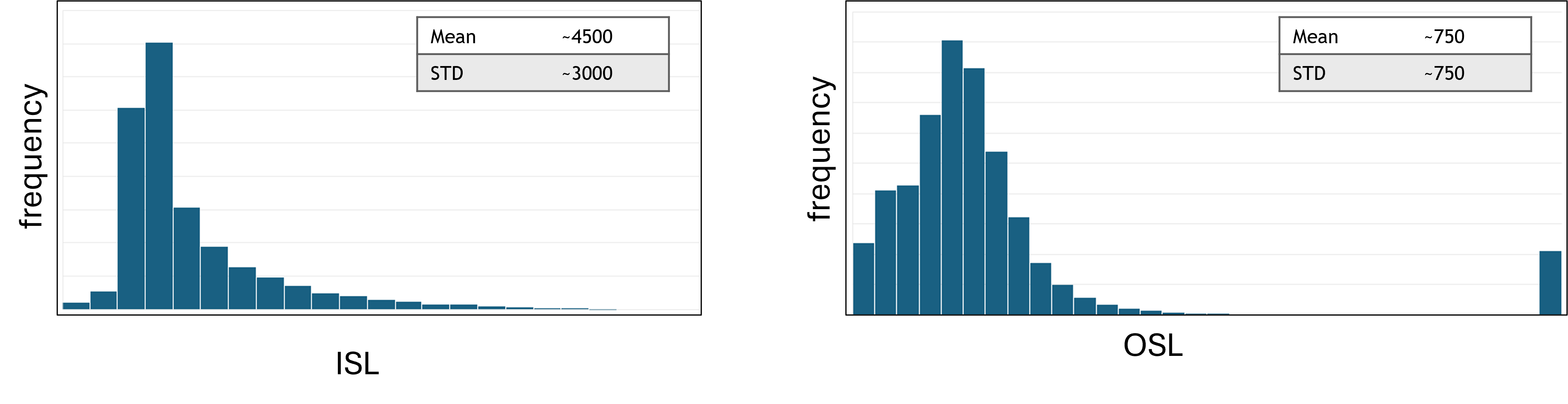}
    \caption{Example distribution of ISL and OSL in dynamic traffic.}
    \label{fig:dynamic_isl_osl_distribution}
\end{figure}

Figure \ref{fig:dynamic_isl_osl_distribution} presents the CDF of input and output sequence lengths (ISL and OSL) from a real-world deployed workload. 
Absolute values have been obfuscated for privacy.

\begin{figure}[htbp]
    \centering
    \includegraphics[width=0.6\textwidth]{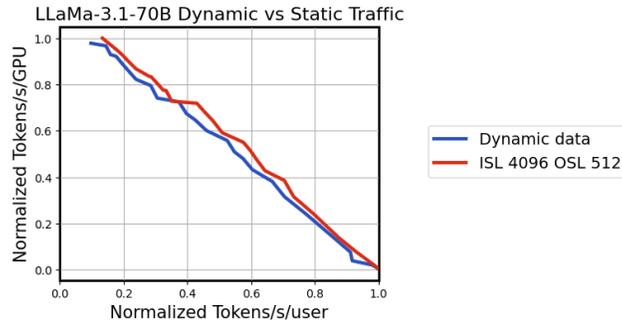}
    \caption{Comparison of Pareto frontiers using dynamic traffic simulation versus P50 approximation.}
    \label{fig:dynamic_vs_static}
\end{figure}

Figure \ref{fig:dynamic_vs_static} shows the resulting Pareto frontiers when this traffic distribution is simulated directly versus
when it is approximated using a simplified approach: the closest power-of-two values to the P50 of ISL and OSL. 
Notably, the approximated frontier closely matches the original, indicating that using P50 ISL and OSL as an approximation provides a reasonable overview of the trends. 

\end{document}